\documentclass[5p,times,twocolumn]{revtex4}
\usepackage{bbm, amsmath, verbatim, graphicx, psfrag, dsfont}
\usepackage[latin1]{inputenc}
\usepackage{color}
\usepackage{verbatim}
\newcommand{\comments}[1]{}
\newcommand{\bra}[1]{\langle #1|}
\newcommand{\ket}[1]{| #1\rangle}
\begin{document}
\title{A single trapped atom in front of an oscillating mirror
\footnote{We devote this paper to our late friend and colleague Krzysztof W\'odkiewicz.}}
\author{A. W. Glaetzle$^{1,3}$}
\author{K. Hammerer$^{1,3}$}
\author{A. J. Daley$^{1,3}$}
\author{R. Blatt$^{2,3}$}
\author{P. Zoller$^{1,3}$}
\address{$\;^1$Institute for Theoretical Physics, University of Innsbruck, A-6020 Innsbruck, Austria\\
$\;^2$Institute for Experimental Physics, University of Innsbruck, A-6020 Innsbruck, Austria\\
$\;^3$Institute for Quantum Optics and Quantum Information of the
Austrian Academy of Sciences, A-6020 Innsbruck, Austria}
\date{\today}

\begin{abstract}
We investigate the Wigner-Weisskopf decay of a two level atom in front of an oscillating mirror. This work builds on and extends previous theoretical and experimental studies of the effects of a static mirror on spontaneous decay and resonance fluorescence. The spontaneously emitted field is inherently non-stationary due to the time-dependent boundary conditions and in order to study its spectral distribution we employ the operational definition of the spectrum of non-stationary light due to the seminal work by Eberly and Wodkiewicz. We find a rich dependence of this spectrum as well as of the effective decay rates and level shifts on the mirror-atom distance and on the amplitude and frequency of oscillations of the mirror. The results presented here provide the basis for future studies of more complex setups, where the motion of the atom and/or the mirror are included as quantum degrees of freedom.

\end{abstract}

\maketitle


\section{Introduction}
Recent experiments with trapped single ions in front of a mirror realize a quantum optical setup where the atom acts as a single photon emitter while the mirror reshapes the electromagnetic environment of the quantized light field \cite{blatt, blatt2,Dubin2007}. Such a half cavity setup allows one to study the interplay between measurement and the wave interference  of the emitted light field introduced by the mirror, as revealed in the spectrum and the second order intensity correlation functions \cite{Morawitz,Cook1987,Alber1992,Dorner}. For a trapped ion homodyne detection of the emitted light allows one to infer the ion motion, and thus implement a quantum feedback scheme to cool the ion \cite{Viktor}. In addition, the light emitted by the ion and scattered by the mirror can act back on the atom with a time delay resulting in a non-Markovian dynamics of spontaneous emission \cite{Cook1987,Alber1992,Dorner} with possible applications for mirror mediated cooling of the ion's motion \cite{Horak2009,Xuereb2009a,Xuereb2009b}.

Below we will extend these discussions by studying a scenario of a single ion in front of an {\it oscillating} mirror. Such a setup is motivated by possible future experiments to combine a mirror attached to a cantilever representing a high-Q quantum oscillator, i.e. opto-nanomechanic system \cite{Marquardt2009}, with a single trapped ion as a composite quantum system. First generation experiments will, however, be certainly in the regime where the motion of the mirror can be treated classically and thus merely constitute a time-dependent boundary condition for the electromagnetic field. For this setting we will investigate the dynamics of spontaneous emission, and in particular the spectrum of emitted light, from a trapped ion in front of an oscillating mirror.

Due to the time-dependent boundary condition the emitted field is non-stationary, even in steady state. In order to calculate the spectrum of the light emitted by the ion we therefore make use of the definition of the \textit{time-dependent physical spectrum of light} due to Eberly and W\'odkiewicz \cite{Eberly1, Eberly2}. In their seminal work the spectrum is defined operationally as the count rate of a photodetector measuring the output field of a tunable filter element, such as a Fabry-Perot cavity.

In the present system the spectrum is determined from the interplay of photon emission and reabsorption processes (as in the static case) along with creation of sideband photons on reflection off the oscillating mirror. The latter process gives rise to interference effects of different spectral components of light, which depend sensitively on the position of the ion as well as on the oscillation frequency and amplitude of the mirror and strongly influence the spectrum of spontaneously emitted photons.

\subsection{The Model}
\begin{figure}[t]
  \centering
  \includegraphics[width=8.4 cm]{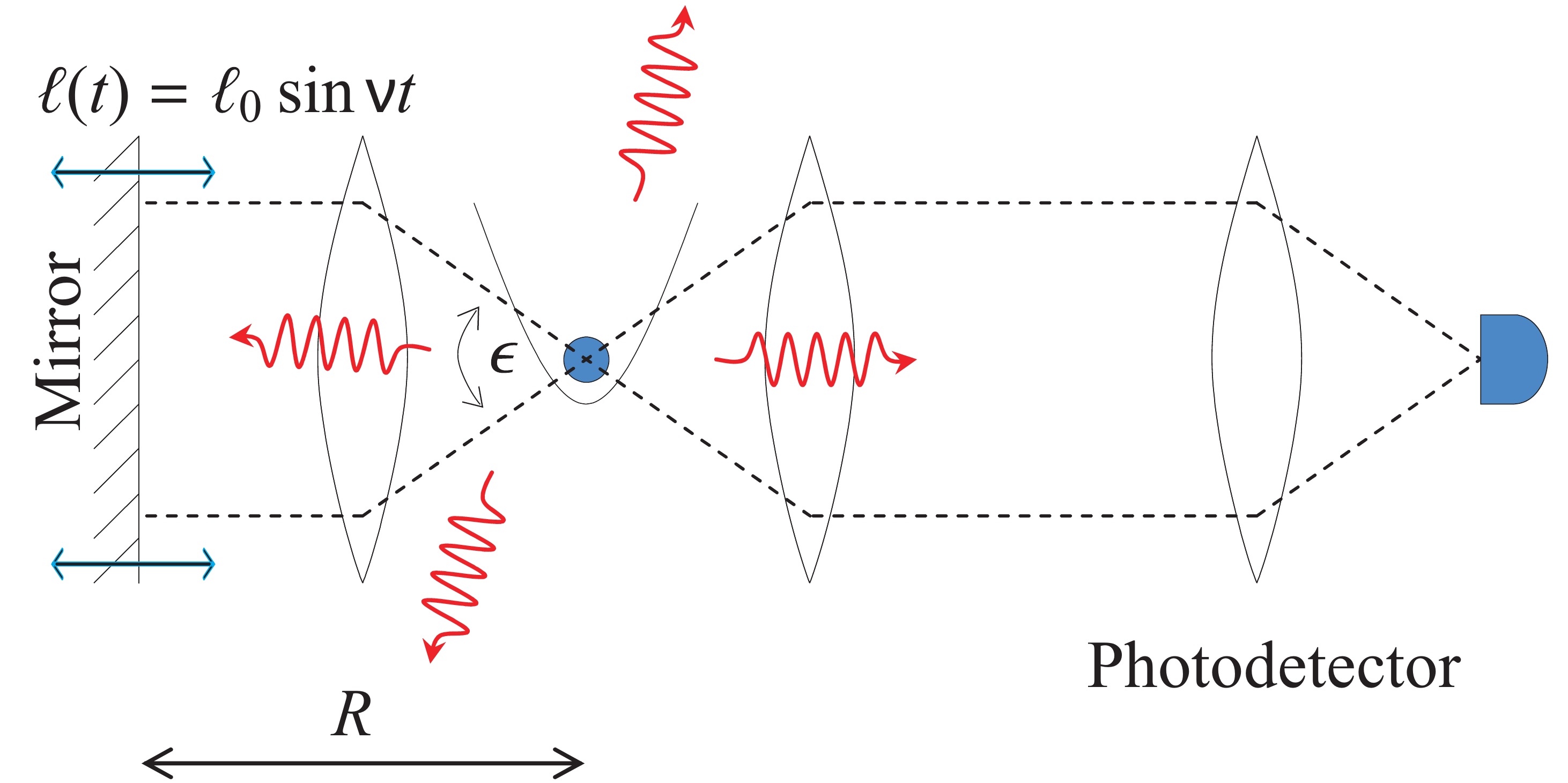}
  \caption{Schematic experimental setup: We consider an atom at position $R$ in
front of a mirror, oscillating on a trajectory $\ell(t)=\ell_0\sin{\nu t}$. Two lenses collimate light, which falls into a solid angle fraction $\epsilon$ and is directed onto the mirror. A photodetector is used to measure the light emitted by the atom.}\label{fig:setup}
\end{figure}
We consider a single trapped atom in front of an oscillating mirror, where the oscillations occur along a prescribed trajectory $\ell(t)=\ell_0\sin{\nu t}$ about the origin. Photons emitted by the atom will be collected in a solid angle $2\pi\epsilon$, cf. Fig.~\ref{fig:setup}, directed towards the oscillating mirror and reflected back onto the atom. These photons are modified by the oscillating mirror before they interact with the atom again. Our goal here is to investigate the resulting modifications to the spontaneous emission rate and the spectrum of fluorescence light. To this end we assume the atom is initially prepared in its excited state and determine the dynamics in a Wigner-Weisskopf approach.

\begin{figure}[t!]
  \centering
  \includegraphics[width=8 cm]{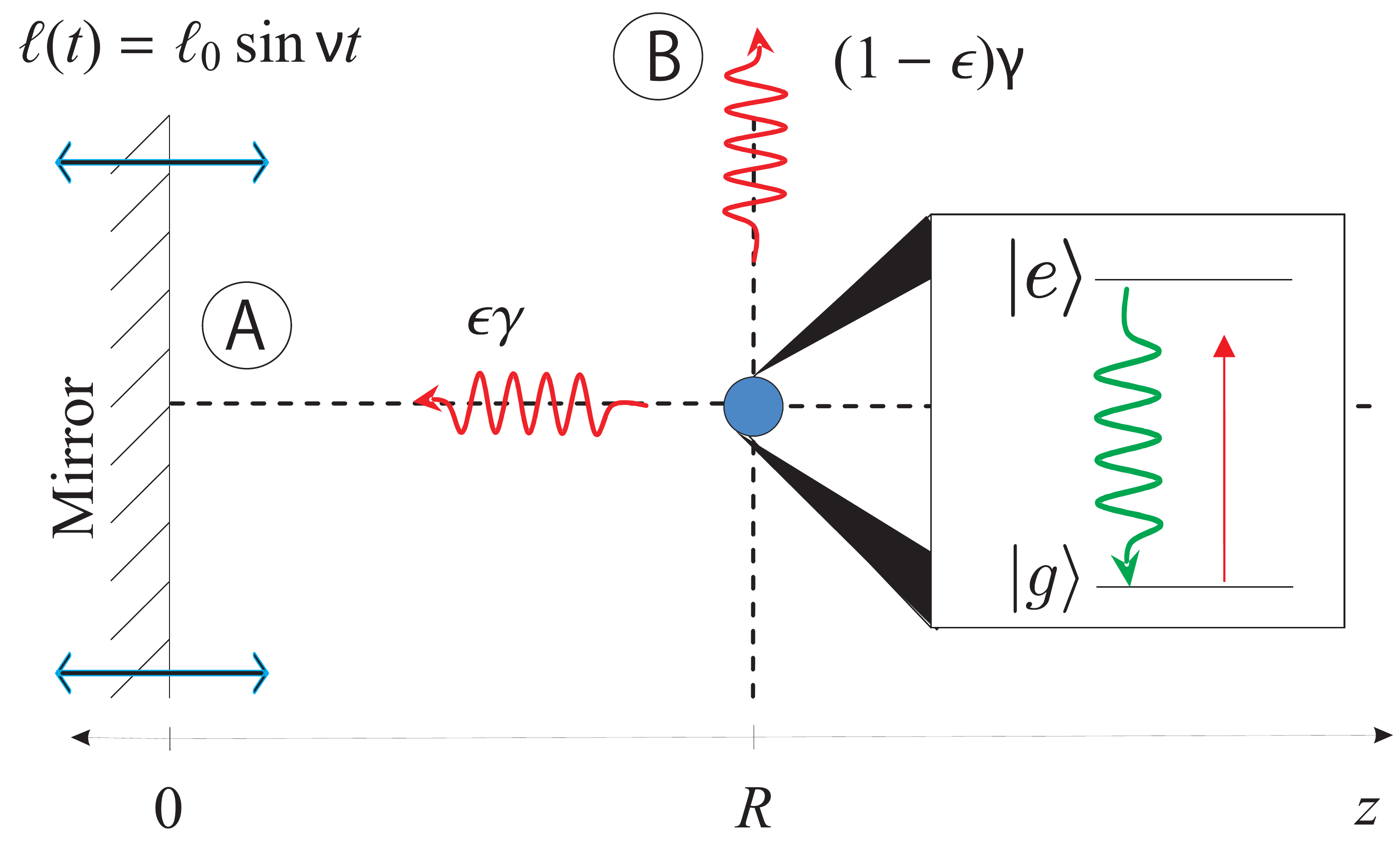}
  \caption{We model the system as two one-dimensional fields interacting
with a two-level atom. Field A describes photons which fall into a solid angle fraction $\epsilon$ onto the mirror and can therefore be described by adiabatically changing standing wave modes. Field B describes photons which are emitted in all other directions and can therefore be described by traveling wave field modes. }\label{fig:ideal}
\end{figure}

We assume that the mirror is oscillating so that its velocity $\ell_0\nu\equiv \mathrm{v}_\mathrm{mirror}\ll c$. This allows us to neglect relativistic corrections and to restrict our model to quasi-static field modes only. \comments{(see Appendix \ref{sec:field}).} Due to the geometry the atom is coupled to two fields. If the emitted photon falls into the solid angle $2\pi\epsilon$, it has to fulfill the time-dependent boundary condition $\mathbf{E}(x,y,z=\ell(t))=0$ and can therefore be described via adiabatically changing standing wave modes. Emitted photons in all other directions are not affected by the mirror and will be described as traveling wave field modes. In order to simplify our calculations we describe the photon bath with two one-dimensional models, similarly to ref. \cite{Dorner} (see Fig.~\ref{fig:ideal}). Based on these assumptions, we calculate the spontaneous emission rate and the spectrum of fluorescence light.

The atomic Hamiltonian describing a two-level atom with ground state energy $E_g=0$ is
\begin{equation}
H_\mathrm{atom}=\hbar\omega_0\ket{e}\bra{e},
\end{equation}
where $\hbar\omega_0$ is the energy difference between the excited and ground states.
We denote the field propagating along the mirror-atom-axis field A and write the Hamiltonian and the positive frequency part of the electric field operator as \comments{(see appendix \ref{sec:field})}
\begin{subequations}
\begin{align}
E_A^{(+)}(z)&=i\int_0^\infty \mathrm{d}k \;\alpha(k) \sin k[z-\ell(t)] a(k) \label{eq:efieldA} \\
H_{A}&=\int_0^\infty \mathrm{d}k \; \hbar \omega(k) \; a^\dag(k) a(k) \label{eq:HamA}.
\end{align}
\end{subequations}
Field B, which is not affected by the mirror, is assumed to be perpendicular to the atom-mirror-axis with
\begin{subequations}
\begin{align}
E_B^{(+)}(x)&=i\int_{-\infty}^\infty \mathrm{d}k \;\beta(k) e^{ikx} b(k)\label{eq:efieldB} \\
H_{B}&=\int_{-\infty}^\infty \mathrm{d}k\; \hbar \omega(k)  \; b^\dag(k) b(k)\label{eq:HamB}.
\end{align}
\end{subequations}
The bosonic operators $a(k)$, $a^\dag(k)$ and $b(k)$, $b^\dag(k)$ annihilate or create a photon in the $k$-th mode of the respective fields, and the factors $\alpha(k)$ and $\beta(k)$ are assumed to be approximately constant in a frequency band of relevance around the frequency $\omega_0$ (narrow band approximation).

Under these assumptions the interaction Hamiltonian describing the interaction of the two photon fields with the atom can be written in rotating wave approximation as
\begin{equation}\label{eq:intham}
H_{\rm int}=-\mu\left( E_A^{(+)}(R)+E_B^{(+)}(0)\right)\ket e \bra g +\mathrm{h.c.}.
\end{equation}
Here we use the electric field operators defined in equations (\ref{eq:efieldA}) and (\ref{eq:efieldB}), , and assume that the electric dipole operator can be written as $\hat \mu=\mu(|\text{e}\rangle\langle\text{g}|+|\text{g}\rangle\langle\text{e}|)$. The Hamiltonian for the transverse electromagnetic field becomes $$H_\mathrm{field}=H_A+H_B,$$ with $H_A$ and $H_B$ the free field Hamiltonians for the two baths defined in (\ref{eq:HamA}) and (\ref{eq:HamB}).

\section{Spontaneous emission rate}\label{sec:sponem}
In this section we investigate the spontaneous emission rate of an atom
initially prepared in the excited state at position $z=R$ in front of
the oscillating mirror with no photons in
the field, $\ket{\psi(0)}=\ket e \otimes \ket{\mathrm{vac}}$, and no laser driving. Therefore we make a Wigner-Weisskopf-type ansatz
\begin{align}
\ket{\psi(t)}&=\;c(t)\ket e \otimes \ket{\mathrm{vac}}\nonumber\\
&\quad+\ket g \otimes \int\mathrm{d}k c_A(k,t) a^\dag(k) \ket{\mathrm{vac}} \\
&\quad+\ket g \otimes \int \mathrm{d}k c_B(k,t) b^\dag(k) \ket{\mathrm{vac}},\nonumber
\end{align}
where $|c(t)|^2$ is the probability to find the atom at time $t$ in the excited state and no photons in the fields, and $|c_A(k,t)|^2$ and $|c_B(k,t)|^2$ are the probabilities to find the atom in its ground state and a photon with wavenumber $k$ in the bath A or B with their corresponding mode functions, respectively. Initially the atom is prepared in its excited state with no photons in the fields and therefore $c(0)=1$ and $c_A(k,0)=c_B(k,0)=0$. In the rotating wave approximation, this ansatz contains all accessible states under time evolution \cite{milonni}.
From the total Hamiltonian $H=H_{\mathrm{atom}}+H_{\mathrm{field}}+H_{\mathrm{int}}$ we derive a system of coupled differential equations for the slowly varying amplitudes $\tilde
c(t)\equiv c(t) e^{i\omega_0 t}$ and $\tilde c_{A,B}(k,t)\equiv c_{A,B}(k,t)
e^{i\omega(k) t}$,
\begin{subequations}
\begin{align}
\dot{\tilde{c}}(t)&= -g \int \mathrm{d}k \;\sin k[R-\ell(t)]e^{-i(\omega(k)-\omega_0)t}\tilde{c}_A(k,t)\nonumber\\
&\quad- h \int \mathrm{d}k\; e^{-i(\omega(k)-\omega_0)t}\tilde{c}_B(k,t),\label{eq:diffct}\\
\dot{\tilde{c}}_A(k,t)&=g \sin k[R-\ell(t)] e^{i(\omega(k)-\omega_0)t} \tilde{c}(t),\label{eq:diffcAt}\\
\dot{\tilde{c}}_B(k,t)&=h \,e^{i(\omega(k)-\omega_0)t} \tilde{c}(t),\label{eq:diffcBt}
\end{align}
\end{subequations}
where we have defined $g\equiv \alpha(k_0) \mu/\hbar$ and $h\equiv \beta(k_0) \mu/\hbar$ following the notation of \cite{Dorner}. In contrast to \cite{Dorner} the standing wave mode functions are now explicitly time-dependent due to the oscillation of the mirror.

In order to solve this infinite system of coupled differential equations we assume a separation of timescales
\begin{equation}
\max{\{\gamma,k_0\ell_0\nu\}}\ll\min{\left\{c/\ell_0, \omega_0\right\}},
\end{equation}
where $\ell_0/c$ is the time the light needs to travel the distance of the mirror's oscillation amplitude. In other words, the spontaneous decay rate $\gamma$ and the Doppler shift $k_0\ell_0\nu$ have to be much smaller than the optical frequency and $\ell_0/c$. Under these assumptions we find (see Appendix \ref{app:adiabaticapprox}) for the excited state amplitude
\begin{equation}\label{eq:nonmarkovpop}
\dot{\tilde{c}}(t)=-\frac{\gamma}{2}\left[c(t)-\epsilon f_+^*(t,t-\tau)e^{i\omega_0\tau}\tilde c(t-\tau)\Theta(t-\tau)\right]
\end{equation}
with
\begin{equation}\label{fplus}
f_+(t,t')\equiv e^{ik_0[\ell(t)+\ell(t')]},
\end{equation}
where we have introduced $\tau\equiv 2R/c$, the mean round trip time from the atom to the mirror and back. As done in ref.~\cite{Dorner}, we have also split up the spontaneous decay rate $\gamma$ into two parts, $\epsilon\gamma\equiv \pi g^2/c$ and $(1-\epsilon)\gamma\equiv 2 \pi h^2/c$, corresponding to the two channels. The first term
on the right hand side of Eq.~\eqref{eq:nonmarkovpop} is the usual free space exponential decay, while the second
term represents the effect of the modified reflected radiation on
the atom which was emitted at time $\tau$ before it interacts
again with the atom. Thus, the retarded argument of the
excited state amplitude directly indicates the memory
effects which are inherent in the system. Furthermore,
the second term is weighted with the factor $\epsilon$ corresponding to the fraction of emitted light that is reflected. The factor $f_+(t,t')$ accounts for the modification due to the oscillating mirror, and would be 1 in the case of a static mirror.

Using the Laplace transform and iteration in $\epsilon$, the solution of Eq.~\eqref{eq:nonmarkovpop} up to first order in $\epsilon$ is
\begin{equation}\begin{split}\label{excitedstate}
  \tilde c(t)=&e^{-\gamma t/2}+\epsilon \frac{\gamma}{2}\; e^{i\omega_0\tau}\Pi_0 e^{-\frac{1}{2} \gamma  (t-\tau )}(t-\tau ) \Theta(t-\tau )\\
  +&\epsilon \frac{\gamma}{2}\; e^{i\omega_0\tau}\sum_{\substack{m=-\infty\\m\neq 0}}^\infty\Pi_m  \left(\frac{e^{-i m \nu  \tau }-e^{-i m \nu  t }}{im\nu }\right)e^{-\frac{1}{2} \gamma  (t-\tau )} \Theta(t-\tau ),
  \end{split}\end{equation}
where we have introduced
\begin{equation}\begin{split}\label{Pifactor}
    \Pi_m(k_0\ell_0,\nu\tau)\equiv\sum_{n=-\infty}^\infty\mathrm{J}_n(k_0\ell_0) \mathrm{J}_{m-n}(k_0\ell_0)  \;e^{in\nu \tau}.
\end{split}\end{equation}
\begin{figure}[tb]
  \centering
  \includegraphics[width=8cm]{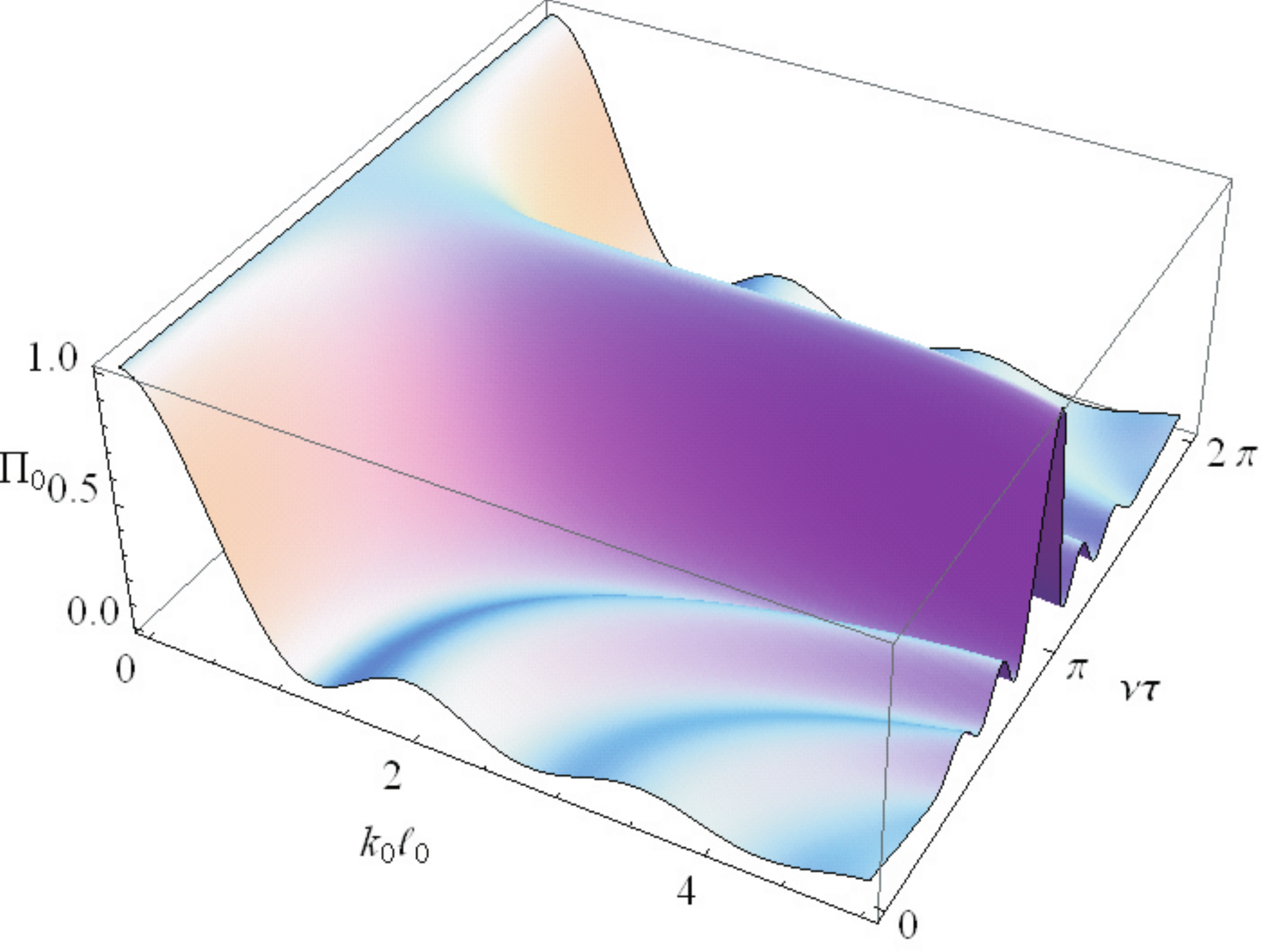}
  \includegraphics[width=8cm]{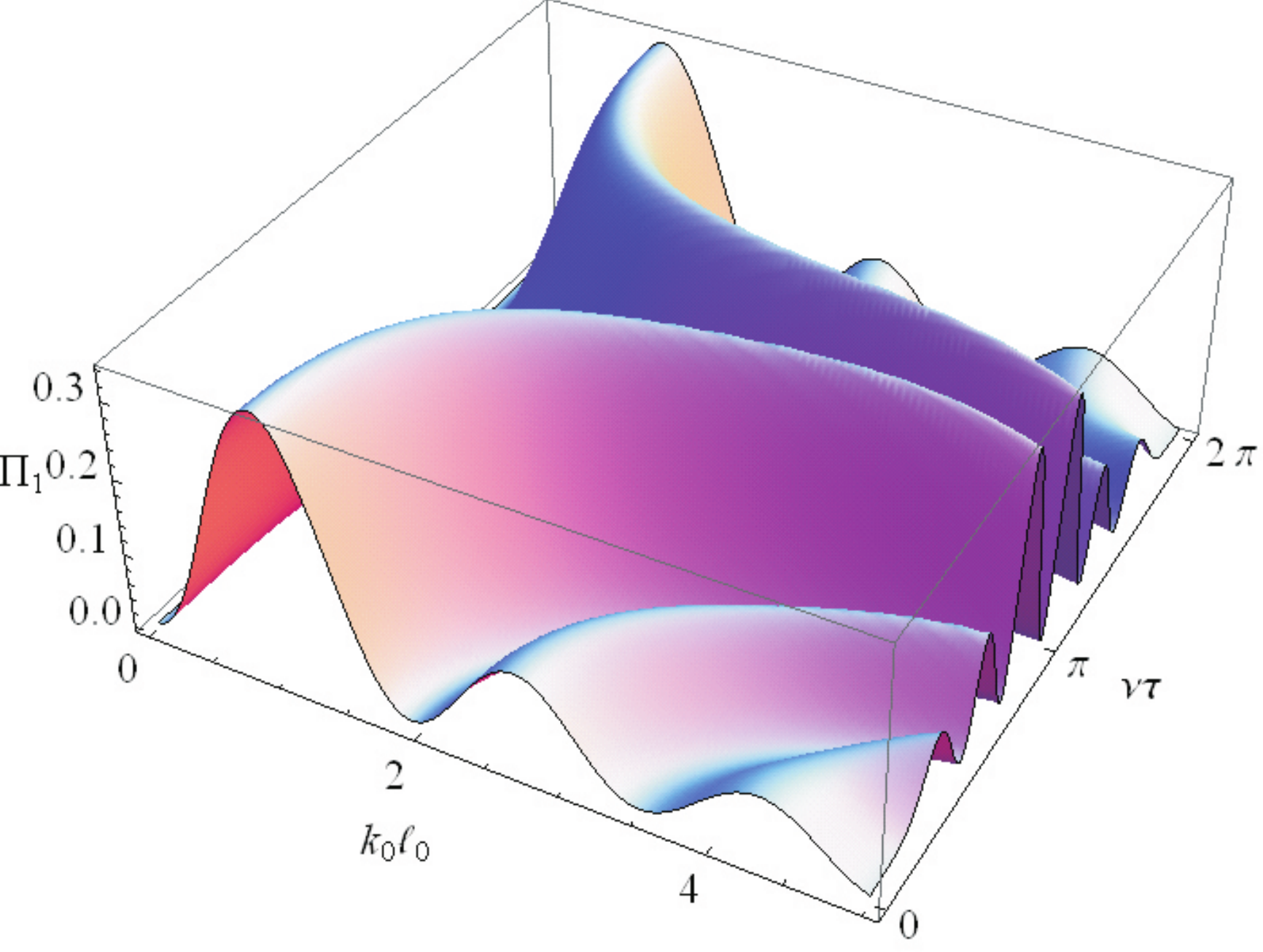}
  \caption{The factors
$\Pi_m(k_0, \nu\tau)$ comprise the effects of the oscillating mirror. In the upper figure $\Pi_0$ and in the lower figure $\Pi_1$ is shown. In the static case, which is $\ell_0=0$ and $\nu=0$, and also in the case $\nu\tau=\pi$ we find $\Pi_m=\delta_{m0}$. }\label{fig:phase2d}
\end{figure}
The first term on the right hand side of Eq.~\eqref{excitedstate} is the familiar Wigner-Weisskopf decay of the probability amplitude of the excited state, while the other terms, proportional to $\epsilon$, give the amplitudes for coherent reexcitation of the atom. The factors $\Pi_m(k_0\ell_0,\nu\tau)$ comprise the effects of mirror oscillations. In the static case, $\nu=0,~\ell_0=0$, one finds $\Pi_m(0,0)=\delta_{0m}$ and correctly recovers from the first line of \eqref{excitedstate} the corresponding result of \cite{Dorner} (i.e. the $n=0$ and $n=1$ terms of Eq.~(7) therein). For an oscillating mirror the factors $\Pi_m(k_0\ell_0,\nu\tau)$ are shown in Fig.~\ref{fig:phase2d} for $m=0$ and $1$.

Remarkably, also for a situation where the condition $\nu\tau=\pi$ (modulo $2\pi$) holds, we find that $\Pi_m(k_0\ell_0,\pi)=\delta_{0m}$, so that the dynamics are identical to the static case.
This phenomenon arises due to interference between sideband photons with frequency shifts $\pm n\nu$ ($n$ integer). They accumulate a phase of $\pm\pi/2$ when they are traveling from the mirror to the atom, and thus they interfere destructively at the position of the atom.

\begin{figure}[tb]
  \centering
  \includegraphics[width=7cm]{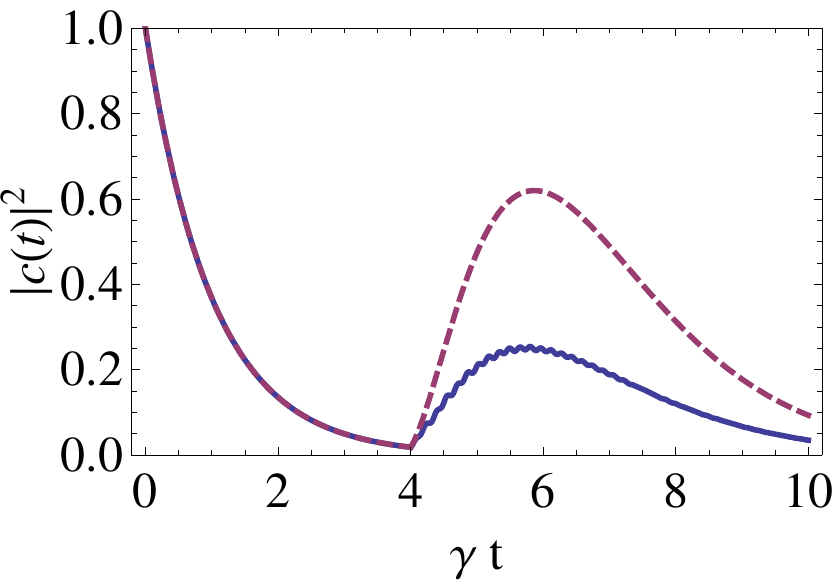}
  \caption{An example for the excited state probability, $|c(t)|^2$ for $\nu = 20, \gamma = 1, \tau = 4, \epsilon = 1, k_0\ell_0 = 1, \omega_0\tau = 0$. The dashed line corresponds to the static mirror and the solid line to the oscillating mirror. At $t=\tau$ the atom is reexcited due to the reflected photons. In the case of an oscillating mirror the photons scattered in sidebands cause a oscillating modulation on the reexcitation peak. }\label{fig:expop}
\end{figure}

The second term in the first line of (\ref{excitedstate}), which is proportional to $(t-\tau)$, represents reexcitation of the atom after a time $\tau$. The factor $\Pi_0$ varies between 0 and 1 (see Fig.~\ref{fig:phase2d}) and therefore, the reexcitaion peak is smaller than in the static case. The second line of (\ref{excitedstate}) adds an oscillating modulation to the reexcitation peak (see Fig.~\ref{fig:expop}), with the latter term being caused by sideband photons interacting with the atom. Fig.~\ref{fig:expop} shows a typical example of the excited state population in front of an oscillating mirror and compares it with an atom in front of a static mirror. The reexcitation peak is smaller due to the $\Pi_0(k_0\ell_0,\nu\tau)$ factor and carries a small modulation which is damped out.

\subsection{Markovian limit}
\begin{figure}[t]
 \centering
 \includegraphics[width=4.3cm]{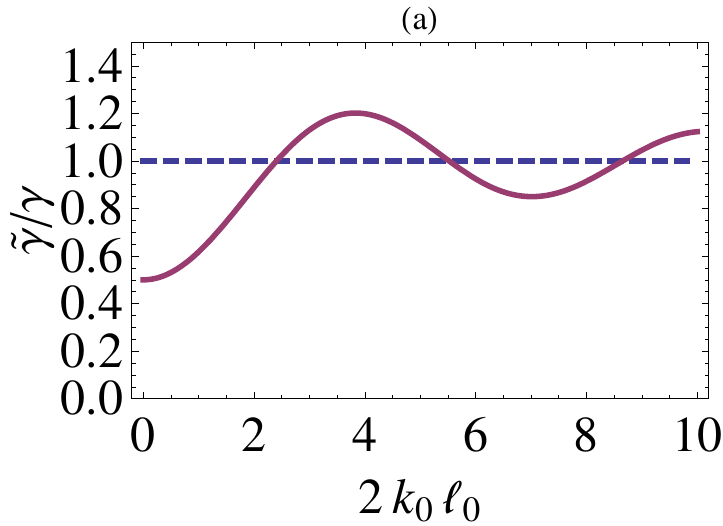}
 \includegraphics[width=4.3cm]{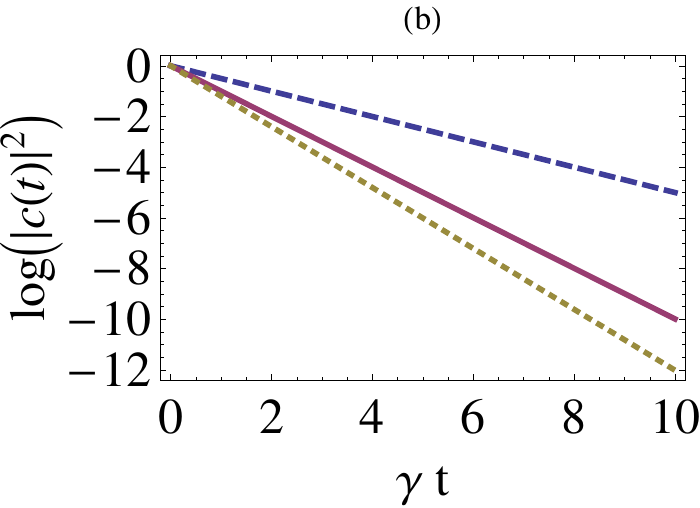}
 \caption{Figure (a) shows the dependence of the decay rate (\ref{modgam}) on the mirror's oscillation amplitude. For $\tilde\gamma/\gamma>1$ the decay is enhanced while for $\tilde\gamma/\gamma<1$ it is inhibited. At the intersections with the dashed line ($\tilde\gamma=\gamma$) the decay rate is not modified in comparison to the free space behavior. Figure (b) shows three examples of the excited state amplitude as a function of time for different oscillation amplitudes of the mirror: $\ell_0=0$ (dashed line), $\ell_0=0.2\lambda_0$ (solid line) and $\ell_0=0.3\lambda_0$ (dotted line) at $\omega_0\tau=2n\pi$ (node). }\label{fig:gammamod}
\end{figure}

If the lifetime of the atom and the time for one period of oscillation of the mirror are both large as compared to the round-trip time $\tau$, that is if
$\gamma\tau\ll1 \quad \text{and} \quad \nu\tau\ll1$,
we can neglect retardation effects and let $\tau\rightarrow0^+$ in the arguments of the step and trigonometric functions in Eq.~(\ref{eq:nonmarkovpop}). In this limit we find
\begin{equation}\label{eq:markovpop}
\dot{\tilde{c}}(t)=-\frac{\gamma}{2}\left[1-\epsilon e^{i\omega_0\tau}e^{-i2k_0\ell(t)}\right]\tilde c(t)
\end{equation}
with an effective, time-dependent, round trip time $\tau_{\rm{eff}}(t)=2(R-\ell(t))/c$.
Eq.~(\ref{eq:markovpop}) can then be solved up to all orders in $\epsilon$ and we find
\begin{multline*}
\tilde c(t)=\exp\left\{-\frac{\gamma}{2}\left[1-\epsilon e^{i\omega_0\tau} \mathrm{J}_0(2k_0\ell_0)\right]t\right\}\\
\times\exp\left\{-\epsilon\frac{\gamma}{2} e^{i\omega_0\tau}\sum_{n\neq 0} \mathrm{J}_n(2k_0\ell_0)\frac{e^{-in\nu t}-1}{in\nu} \right\}.
\end{multline*}
Introducing modified position-dependent decay rate and level spacing similar to \cite{Dorner}
\begin{subequations}\label{moddet}
\begin{eqnarray}
&\tilde\gamma\equiv\gamma\left(1-\epsilon\;\mathrm{J}_0(2k_0\ell_0) \cos{\omega_0\tau}\right),\label{modgam}\\
&\tilde\omega_0\equiv\omega-\epsilon\frac{\gamma}{2}\;\mathrm{J}_0(2k_0\ell_0) \sin{\omega_0\tau},\label{moddet}
\end{eqnarray}
\end{subequations}
we find (in the non-rotating frame)
\begin{equation}\label{popexmarkov}
c(t)= e^{-i\tilde\omega_0 t}e^{-\tilde\gamma t/2}\exp\left\{-\epsilon\frac{\gamma}{2} e^{i\omega_0\tau}\sum_{n\neq 0} \mathrm{J}_n(2k_0\ell_0)\frac{e^{-in\nu t}-1}{in\nu} \right\}.
\end{equation}
The amplitude of the atom in the excited state decays exponentially at a modified rate $\tilde\gamma$, defined in (\ref{moddet}), and carries an oscillating modulation. The amplitude of this modulation decreases when the oscillation frequency of the mirror increases. Thus, in the resolved sideband regime, $\gamma\ll\nu$, the modification due to this term is small. As in the case of a static mirror, the effective level shift (\ref{moddet}) and decay rate (\ref{modgam}) depend on the position of the atom in front of the mirror. The mirror's oscillation adds a factor $\mathrm{J}_0(2k_0\ell_0)$ to this position dependence. For a large amplitude of oscillation $k_0\ell_0\gg 1$ we recover the free space values $\tilde\gamma=\gamma$ and $\tilde\omega=\omega$, as in this case one can no longer consider the atom to be located exactly at e.g. a node or antinode. On top of that, free space behavior is also recovered at the nodes of the zeroth-order Bessel function, because the reflection into the carrier frequency $\omega_0$ is suppressed. Figure \ref{fig:gammamod} shows the dependence of the decay rate (\ref{modgam}) and the excited state amplitude \eqref{popexmarkov} on the amplitude of the mirror oscillation. For a static mirror, $\ell_0=0$ or $\nu=0$, we rediscover Eq.~(10) obtained in \cite{Dorner}, where at a node of the standing wave spontaneous decay is inhibited while at an antinode it is enhanced.

\section{Spectrum of the Field}\label{sec:spectrum}

\subsection{Photon population in channel B}
In order to obtain the photon population in channel B (perpendicular to atom-mirror-axis), we have to solve equation (\ref{eq:diffcBt})
with the excited state solution (\ref{excitedstate}). For simplicity, we split the probability amplitude $c_B(k,t)$ into two parts $c_B(k,t)=c_B^{(0)}(k,t)+c_B^{(1)}(k,t)$. In the steady state, the second line of (\ref{excitedstate}) gives rise to the steady state population
\begin{equation}\begin{split}\label{0photonpopB}
c_B^{(0)}(k)=\frac{h}{\gamma /2-i(\omega-\omega_0)}
+\epsilon h \frac{\gamma}{2}\Pi_0\frac{e^{i\omega\tau}} {[\gamma/2 -i (\omega-\omega_0)]^2}.
\end{split}\end{equation}
The first term is the usual free space behavior of a two-level system and the second term is the modification due to the mirror, discussed in \cite{Dorner}. For an oscillating mirror this term is suppressed by the factor $\Pi_0(k_0\ell_0,\nu\tau)$. The second line of (\ref{excitedstate}) gives in the steady state
\begin{multline}\label{1photonpopB}
c_B^{(1)}(k)=h \epsilon\left(\frac{\gamma}{\nu}\right)\sum_{m\neq0}\frac{\Pi_m(k_0\ell_0,\nu\tau)}{2im}\\
\times\,\Bigg[\;\frac{e^{i(\omega-m\nu)\tau}}{\gamma/2-(i\omega-\omega_0-m\nu)}
-\frac{e^{i(\omega-m\nu)\tau}}{\gamma/2-i(\omega-\omega_0)}\Bigg].
\end{multline}
The first term in the square brackets describes sidebands centered at multiples of the oscillation frequency of the mirror, $\nu$, with a width $\gamma/2$. The second term is a modification of the carrier peak. To resolve the sidebands we need $\gamma\ll\nu$, but the prefactor is proportional to $\gamma/\nu$ and thus the height becomes small.
\begin{figure}[tb!]
  \centering
  \includegraphics[width=8 cm]{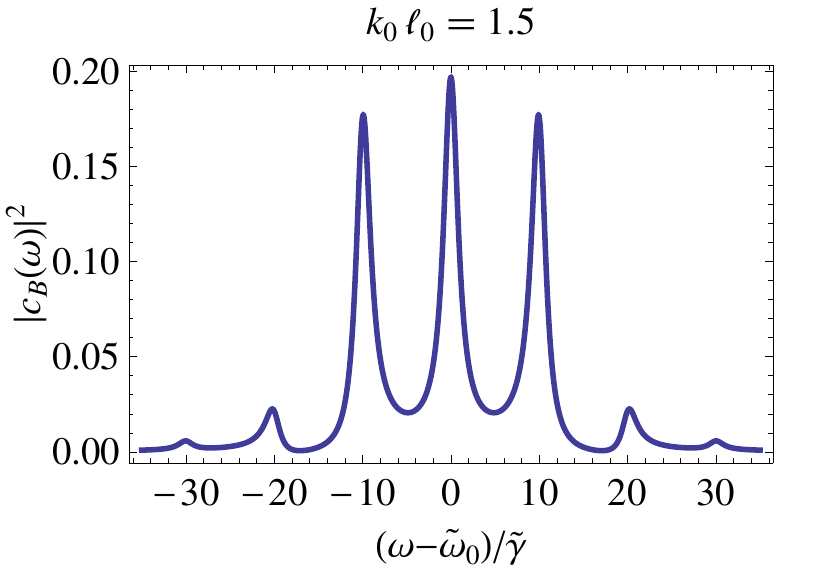}
  \includegraphics[width=8 cm]{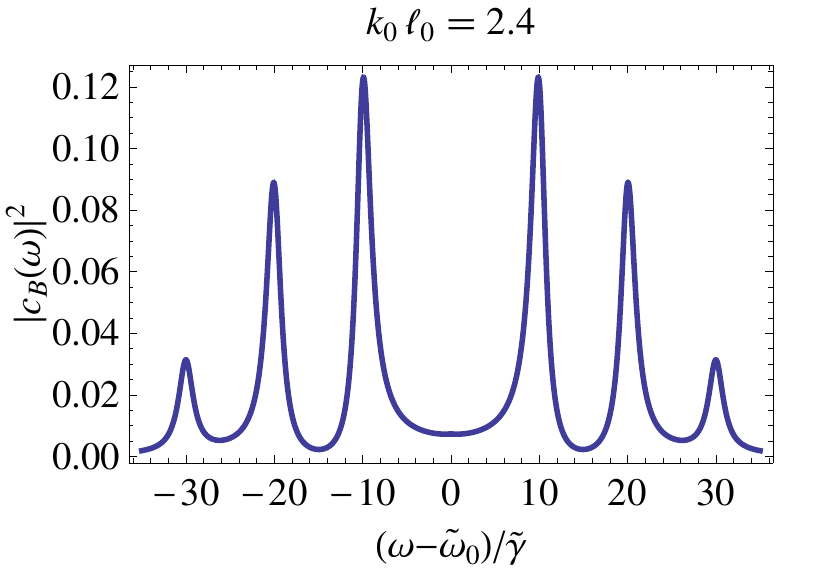}
  \caption{Amplitude dependence of the photon population: In the upper panel the argument of the Bessel functions, $k_0\ell_0$, is chosen so that the peaks decrease with increasing order. In the lower panel the argument of the Bessel functions is chosen so that the zeroth order Bessel function has a node. Therefore the central peak is suppressed.}
  \label{specamplitude}
\end{figure}
\begin{figure}[tb!]
  \centering
  \includegraphics[width=8.4 cm]{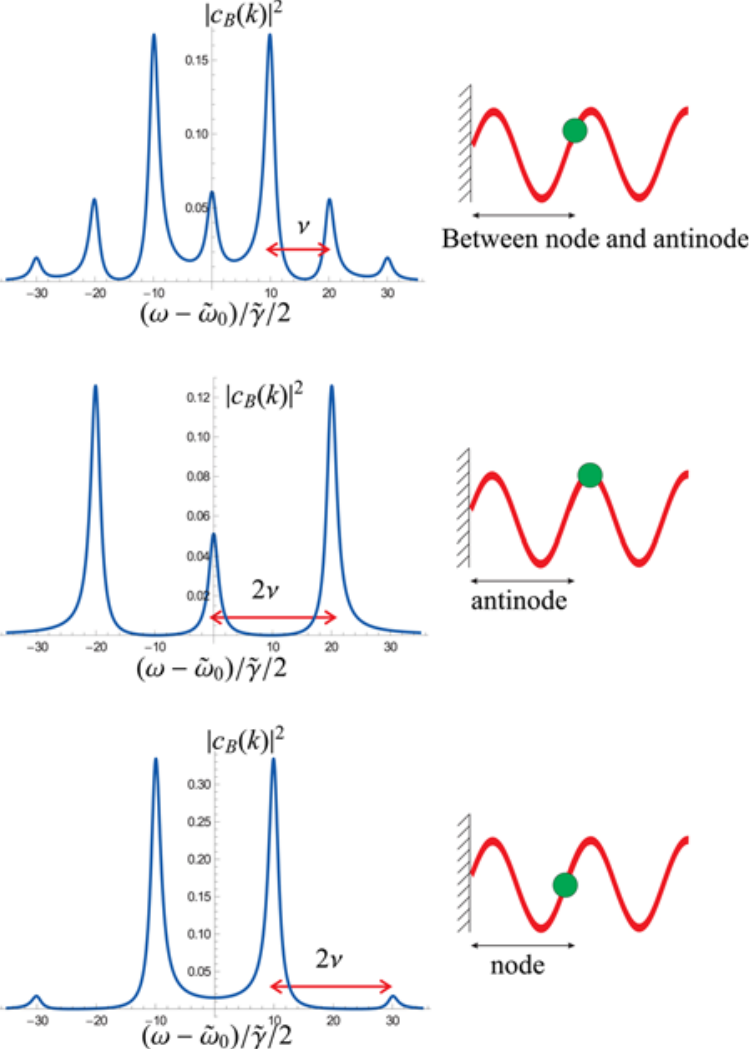}
   \caption{These three figures show the dependence of the photon population on the position of the atom. If the atom is located between a node and an antinode of the resonance frequency $\omega_0$ (upper figure) all peaks are visible. If the atom is located at an antinode (middle figure) only the even peaks are visible, and if the atom is located at a node (lower figure) only the odd peaks are visible.}
  \label{specposition}
\end{figure}

\subsection{Photon population in channel A}
In this section we want to calculate the photon population in the atom-mirror-channel in the \textit{Markovian limit}. As we have seen in the previous section the sidebands due to the modification of the excited state population are proportional to $\gamma/\nu$ and therefore very small in the resolved sideband regime $\gamma\ll\nu$. We therefore neglect this effect and insert only the first two exponentials of (\ref{popexmarkov}) in (\ref{eq:diffcAt}). Expanding the time-dependent mode functions with the
Jacobi-Anger-identity \cite{jacobianger}
$$e^{i z \cos \theta}=\sum_{n=-\infty}^{\infty} i^n\, \mathrm{J}_n(z)\, e^{i n \theta} $$
and solving the resulting differential equation we find
\begin{equation}\label{multimode}
{c}_A(k,t)=g\sum_{n=-\infty}^{\infty}A_n(k,R)\, \frac{e^{-i(n\nu-\omega-\tilde\omega_0)t}e^{-\tilde\gamma t/2}-1}{i(\omega-\tilde\omega_0-n\nu)-\tilde\gamma/2},
\end{equation}
with
\begin{equation}\label{An}
A_n(k,x)=\frac{1}{2i}\left(e^{ikx}-(-1)^ne^{-ikx}\right)\mathrm{J}_n(k_0\ell_0),
\end{equation}
and the modified spontaneous decay rate and detuning introduced in (\ref{moddet}) and (\ref{moddet}).
For $t\rightarrow\infty$ Eq.~(\ref{multimode}) simplifies to
\begin{align*}
\tilde{c}_A(k)=-g  &\left\{\sum_{n=-\infty}^\infty
\frac{\sin(k_0R)\textrm{J}_{2n}(k_0\ell_0)}{i(\omega(k)-\tilde\omega_0-2n\nu)-\tilde\gamma/2}\right.\\
+i&\left.\sum_{n=-\infty}^\infty \frac{\cos(k_0R)
\textrm{J}_{2n-1}(k_0\ell_0)}{i(\omega(k)-\tilde\omega_0-(2n+1)\nu)-\tilde\gamma/2}\right\}.
\end{align*}
Since these are Lorentzian functions with a width $\tilde\gamma$ we have
set $k=k_0$ in the argument of the trigonometric functions.
To distinguish the sidebands and the carrier, $\tilde\gamma$ has to
be much smaller than $\nu$. In the case where $\nu\ll \tilde \gamma$, the sidebands overlap. There are two possibilities to change the relative height of the peaks. First, the peaks are proportional to $\textrm{J}_{n}(k_0\ell_0)$, thus by changing the amplitude of the mirror's oscillation the height of the sidebands will change. For example, if the argument of the Bessel functions is chosen so that the zeroth order Bessel function has a node the central peak is completely suppressed (see Fig.~\ref{specamplitude}). Or second, the odd (even) peaks are proportional to $\cos(k_0R)$ ($\sin(k_0R)$), and thus, by changing the position of the atom one can change the height of the even numbered peaks respectively to the odd numbered peaks. For example, if the atom is located between a node and an antinode of the resonance frequency $\omega_0$  all peaks are visible. If the atom is located at an antinode only the even peaks are visible and if the atom is located at a node  only the odd peaks are visible (see Fig.~\ref{specposition}). 

\begin{figure}[tb!]
  \centering
  \includegraphics[width=8 cm]{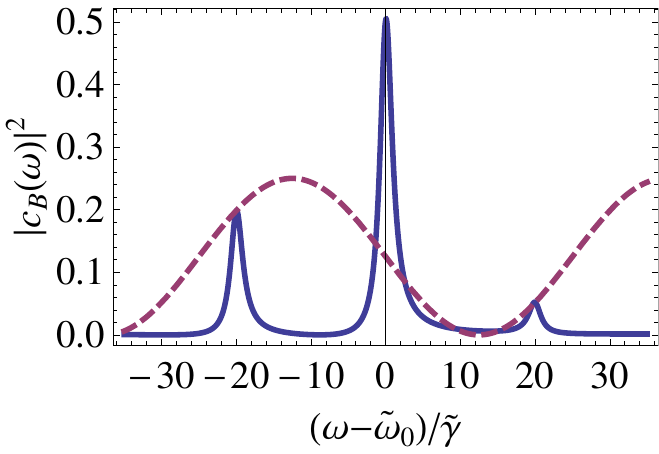}
  \caption{Dependence of the photon population on frequency: In the non-Markovian regime the sidebands become asymmetric due
to the envelope of the trigonometric function.}
  \label{fig:phpopnm}
\end{figure}

In the non-Markovian regime we cannot set $k=k_0$ in the argument of the trigonometric functions and find
\begin{equation*}\begin{split}
  c_A(k)=&-g\sum_{m=-\infty}^\infty \frac{A_m(k,R)}{i(\omega-\omega_0-m\nu)-\gamma /2}\\
  +&\epsilon g \frac{\gamma}{2}\Pi_0\sum_{m=-\infty}^\infty \frac{A_m(k,R)e^{i(\omega-m\nu)\tau}}{[i (\omega-\omega_0-m\nu)-\gamma/2 ]^2}.
  \end{split}\end{equation*}
For $\gamma\tau\leq1$ the sidebands become asymmetric due to the envelope of the trigonometric functions (see Fig.~\ref{fig:phpopnm}).
Therefore, by putting
the atom at the right position, one should, in principle, increase the heating
or cooling rate of a trapped atom, by letting it interact with the blue or red sideband only.

\subsection{Emission Spectrum}
It is important to note that in the present case the squared probability amplitudes $|c_{A(B)}(k)|^2$ do not directly provide the spectrum of the emitted field. This is due to the fact that the electric field will be non-stationary due to the time-dependent boundary conditions even when the probability amplitudes of the system's wave function are in their steady, time-independent state.

\begin{figure}[tb!]
  \centering
  \includegraphics[width=8.4 cm]{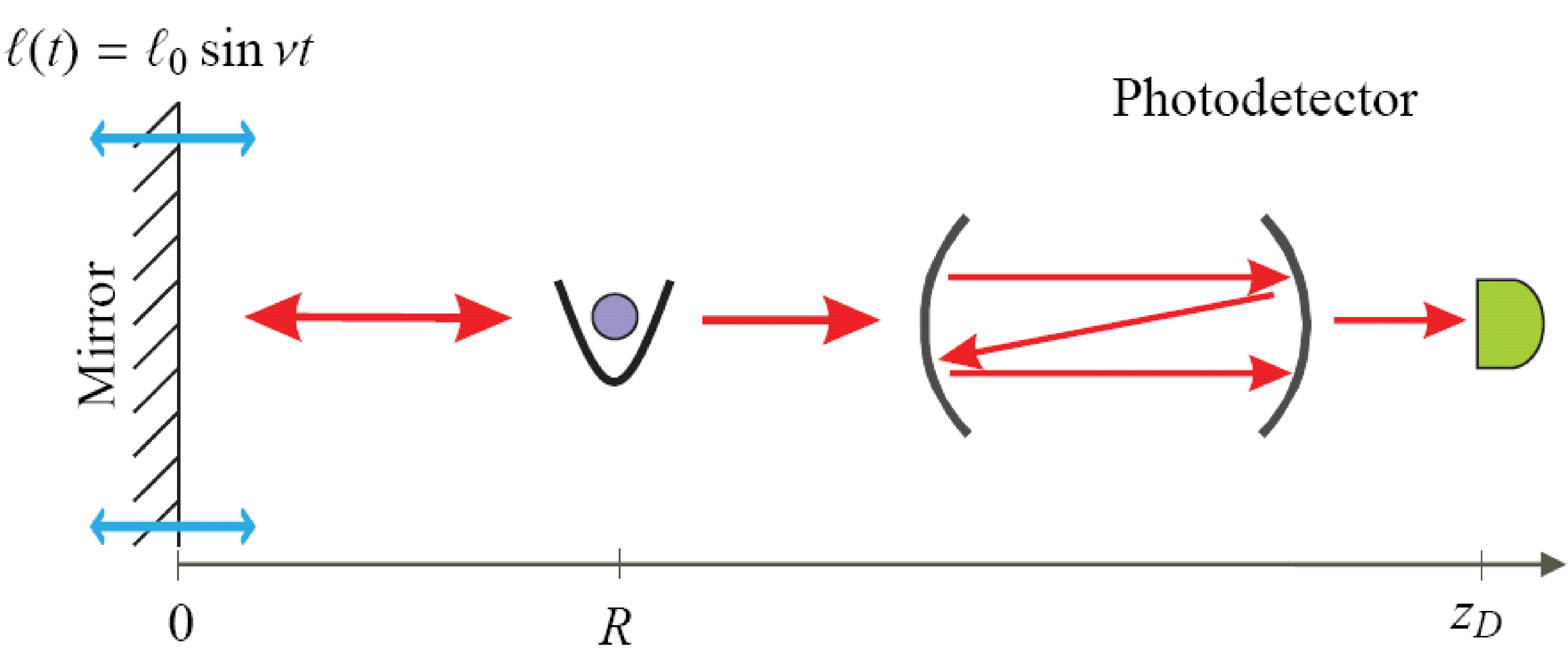}
  \caption{Schematic drawing of the measurement process: A Fabry-Perot cavity is inserted in the mirror channel, and used as a filter to determine the resonant frequency for detection.}
  \label{fig:detector}
\end{figure}

In order to evaluate the spectrum of emitted light we thus make us of the operational definition of the spectrum of non-stationary fields due to Eberly and Wódkiewicz \cite{Eberly1}. This definition considers what is actually detected when a Fabry-Perot filter is inserted in the mirror channel (see Fig.~\ref{fig:detector}) and the energy density
\begin{equation}
w(t)\equiv \langle E^{(-)}_D(z_D,t)E^{(+)}_D(z_D,t)\rangle
\end{equation}
is measured behind it. The detector does not see the entire light field $E(t)$ under study, but a filtered version of it:
\begin{equation}
E^{(+)}_D(t)=\int_{-\infty}^{\infty}\mathrm{d}t' f(t-t') E^{(+)}(t').
\end{equation}
Here $f(t-t')$ is the response function of the filter. A suitable filter would be, for example, a two-sided cavity. The spectral transmission function of such a cavity \cite{colgard} near resonance, $\omega\approx\omega_D$, is
$$\tilde f(\omega)=\frac{\Gamma_D}{\Gamma_D-i(\omega-\omega_D)}$$ with the Fourier transform
\begin{equation}\label{filter}
f(t)=\Gamma_\mathrm{D} \,\Theta(t) \,e^{-(\Gamma_D+i \omega_D)t},
\end{equation}
where $\omega_D$ is the setting frequency and $\Gamma_D$ is the bandwidth of the filter.
In the Wigner-Weisskopf approximation the correlation function factorizes and we obtain
\begin{equation*}
\langle E^{(-)}(z_D,t_1)E^{(+)}(z_D,t_2)\rangle=\mathcal{E}^*(z_D,t_1)\mathcal{E}(z_D,t_2),
\end{equation*}
with the c-number electric field
\begin{equation}\label{cFieldformal}
\mathcal{E}(z,t)=\alpha\int_0^\infty dk \sin{k (z-\ell(t))}\;c_A(k,t).
\end{equation}
Together with (\ref{multimode}) we obtain
\begin{equation}\label{cField}
\begin{split}
\mathcal{E}(R+d,t)=&\frac{\alpha g\pi}{2c}\Big[\Theta\left({\textstyle t-\frac{d}{c}}\right)\;e^{-(i\tilde\omega_0+\tilde\gamma/2)\left(t-\frac{d}{c}\right)}\\
-&\sum_{n,m}\Theta\left({\textstyle t-\frac{2R+d}{c}}\right)\;\mathrm{J}_n(k_0\ell_0)\;\mathrm{J}_m(k_0\ell_0)\\
&\qquad e^{-i(n+m)\nu t}\;e^{-(i\tilde\omega_0+\tilde\gamma/2)\left(t-\frac{2R+d}{c}\right)}\Big],
\end{split}
\end{equation}
where we defined $d=z_D-R>0$ as the distance between the atom and the detector (see Appendix \ref{app:wwspectrum}).
The c-number electric field at the detector consists of an unmodulated part which travels directly from the atom to the detector and a part which travels to the mirror, becomes modulated, and is reflected back to the detector. We see this reflected in the time delays of the two terms, combined with the first term being independent of the mirror oscillation frequency. Note that because we have already assumed that $\tilde\gamma\tau$ was small in (\ref{multimode}), we must take the Markovian limit of (\ref{cField}).
\begin{figure}[tb!]
  \centering
  \includegraphics[width=8.4 cm]{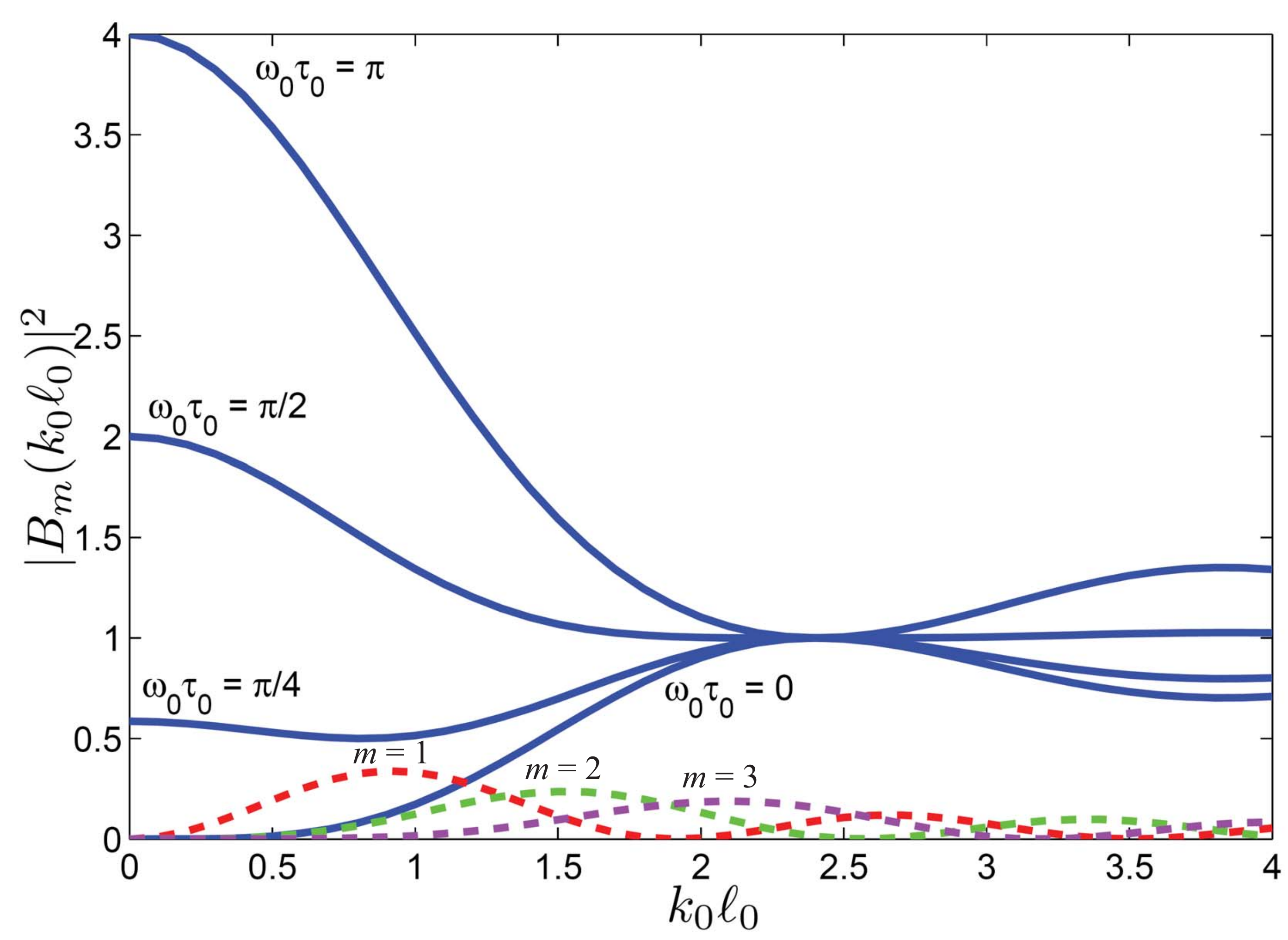}\\
  \caption{Strength, $|B_m(k_0\ell_0)|^2$ for $m=0\ldots3$, of the sidebands at varying oscillation amplitudes for different positions of the atom.}\label{amp}
\end{figure}
We now identify the spectrum as the normalized energy density for different resonant frequencies of the cavity and find in the limit of ideal spectral resolution $\Gamma_D\longrightarrow 0$ and in the Markovian limit $t\gg1/\tilde\gamma, R/c, d/c$
\begin{multline}\label{eq:spectrum}
S(\omega)=\left(\frac{\alpha g\pi}{2c}\right)^2\Big|\frac{B_0(k_0\ell_0)}{i(\omega-\tilde\omega_0)-\tilde\gamma/2}\\
-\sum_{m\neq0}\frac{B_m(k_0\ell_0)}{i(\omega-\tilde\omega_0-m\nu)-\tilde\gamma/2}\Big|^2,
\end{multline}
with
\begin{subequations}
\begin{eqnarray}
&&B_0(k_0\ell_0)\equiv 1-e^{i \omega_0\tau}\;\mathrm{J}_0^2(k_0\ell_0),\label{carrierstrength}\\
&&B_m(k_0\ell_0)\equiv e^{i \omega_0\tau}e^{-im\nu\left(\tau+ \frac{d}{c}\right)}\mathrm{J}_{m}(2k_0\ell_0).\label{sidestrength}
\end{eqnarray}
\end{subequations}
For increasing mirror oscillations, where $\mathrm{J}_0(2k_0\ell_0)\ll1$, the dependence of the carrier on the position of the atom vanishes and $B_0\approx 1$. The oscillating mirror redistributes the intensity of the carrier into the sidebands. This indicates that for large mirror oscillation amplitudes much less light is reflected at the carrier frequency from the mirror than travels directly from the atom to the detector.
In order to resolve the sidebands we assume $\gamma\ll\nu$ and in this case the sidebands add approximately incoherently. Thus we call $|B_m(k_0\ell_0)|^2$ the strength of the $m$-th sideband.

\begin{figure}[tb!]
  \centering
  \includegraphics[width=8 cm]{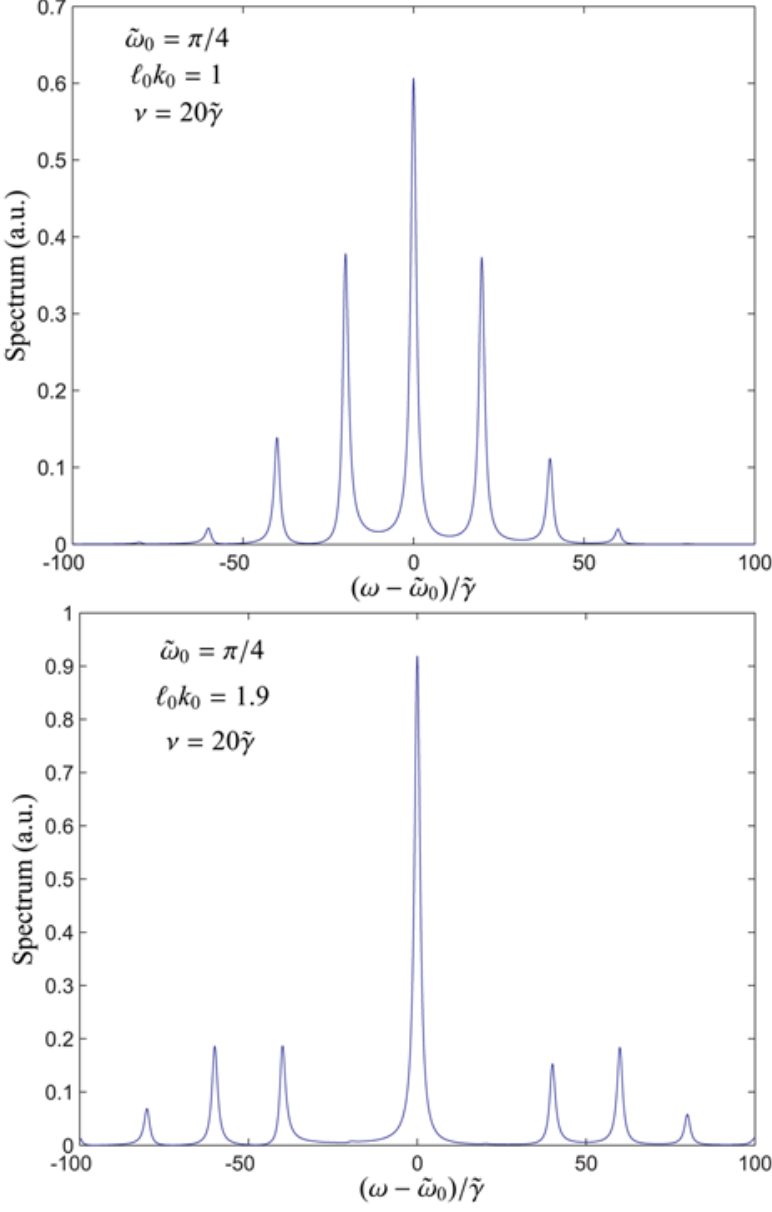}
  \caption{Normalized spectrum for an atom at position $k_0 R=\pi/8$ and for mirror oscillation amplitudes $\ell_0k_0=1$ (upper) and $\ell_0k_0=1.9$ (lower). For $\ell_0k_0=1.9$ the first Bessel function $\mathrm{J}_1(2k_0\ell_0)$ is approximately zero (see figure \ref{amp}) and therefore the first sideband vanishes.}\label{spec4}
\end{figure}

\begin{figure}[tb!]
  \centering
  \includegraphics[width=8 cm]{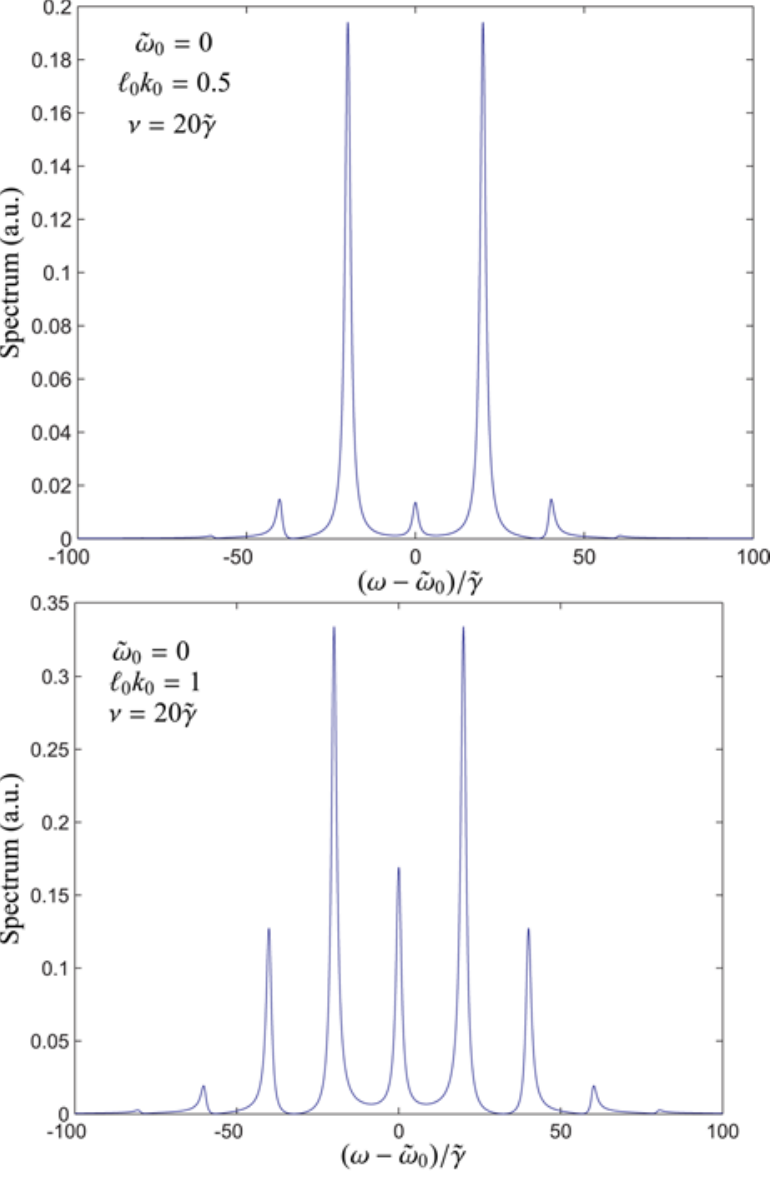}
  \caption{Normalized spectrum for an atom at position $k_0 R=0$ and for oscillation amplitudes $\ell_0k_0=.5$ (upper) and $\ell_0k_0=1$ (lower).}\label{spec5}
\end{figure}

Fig.~\ref{amp} shows the strength of the carrier and the sidebands according to equations (\ref{carrierstrength}) and (\ref{sidestrength}) for different oscillation amplitudes and different positions of the atom. The strength of the carrier depends on the position of the atom in the standing wave. The sidebands pick up a phase when traveling from the mirror to the atom, which leads to interference between sidebands with $\pm\nu$. In the Markovian limit this phase is approximately zero for all relevant sidebands. Fig.~\ref{spec4} shows the normalized spectrum for an atom at position $k_0 R=\pi/8$ and for oscillation amplitudes $\ell_0k_0=1$ and $\ell_0k_0=1.9$. We see that when the argument is chosen to be $2\ell_0k_0=3.8$, the first sideband is suppressed (left figure). The height of the peaks can be compared with Fig.~\ref{amp}. The spacing between the sidebands is $\nu/\tilde\gamma=20$ and the width is normalized to 1. Fig.~\ref{spec5} shows the normalized spectrum for an atom at position $k_0 R=0$ and for oscillation amplitudes $\ell_0k_0=0.5$ and $\ell_0k_0=1$.

\section*{Summary and Outlook}

In this work we have investigated the spontaneous decay and the fluorescence spectrum of an initially excited two-level atom, in front of an oscillating, perfectly reflecting mirror, without any laser excitation. We therefore used the one dimensional model introduced in \cite{Dorner}.

In section \ref{sec:sponem} we have calculated the excited-state amplitude in first order in $\epsilon$ for long atom-mirror distances and up to all orders for small atom-mirror distances. The initially excited atom will decay freely and the emitted photon will fall with a probability $\epsilon$ onto the mirror and gets back-reflected onto the atom. Together with the incoming field a standing wave field will build up in the atom mirror channel which is spatially shifted due to the mirror oscillation. Compared to the static case, an oscillating mirror creates sideband photons with frequency shifts $\pm n \nu$ ($n$ integer). The reflected carrier photons will re-excite the atom after a time $\tau=2R/c$, while the sideband photons will cause an oscillating modulation of the excited state amplitude. When traveling from the mirror to the atom the sideband photons accumulate a phase and will interfere. Thus, for particular atom-mirror distances, the situation is equivalent to an atom in front of a static mirror, when the sideband photons have destructively interfered and the whole energy remains in the carrier. For small atom mirror distances (in the Markovian limit) we find an effective time-dependent round-trip time between the atom and the mirror. In addition to the atomic position dependence inside the standing wave, the decay rate depends also on the amplitude of the mirror's oscillation. The atom can no longer be located at a fixed position in the standing wave because the standing wave is shifted according to the mirror.

In section \ref{sec:spectrum} we first calculated the field amplitude in channel A and B and then the fluorescence spectrum, as it was defined by Eberly and Wodkíewicz. The effect of the oscillating modulation of the excited state amplitude on the photon amplitude in channel B (perpendicular to the atom-mirror axis) is proportional to $\gamma/\nu$ and therefore small in the resolved sideband regime. In channel A (atom-mirror channel) we find that the height of the sideband peaks depends on the mirror's oscillation amplitude and on the position of the atom in the standing wave. Remarkably, in the non-Markovian regime the blue and red sidebands become asymmetric.

Based on the present results we will explore in future work the possibilities to include motional degrees of freedom of the atom, or even of the mirror in the sense of a micro-mechanical oscillator. We expect that the reflected sideband photons will have an appreciable effect on the center of mass motion of the ion, especially in the non-Markovian regime and when the atom is driven continuously by a laser. For such a system cooling or heating effects were predicted by recent theoretical work \cite{Horak2009,Xuereb2009a,Xuereb2009b}. Further possible extensions of this work would be to treat the mirror as a dynamical degree of freedom or to take Casimir-photons, created by the oscillating mirror, into account and investigate their effect on e.g. the center of mass motion of the trapped ion.


\section*{Acknowledgments}
We thank G. H\'{e}tet, L. Slodicka and S. Gerber for stimulating
and helpful discussions. Work at IQOQI and the University of Innsbruck is supported by the Austrian Science Fund through SFB FOQUS, and EU projects.

\appendix

\section{Adiabatic Approximation}\label{app:adiabaticapprox}
In this appendix we calculate the equation of motion for the excited state amplitude, Eq.~\eqref{eq:nonmarkovpop}.
We assume that the atom dominantly couples to a band of frequencies centered at $\omega_0$ with a width $\vartheta$. This cutoff frequency $\vartheta$ has to be much smaller than the optical frequency associated with $\omega_0$. Furthermore, we assume that inside this frequency band, $[\omega_0-\vartheta,\omega_0+\vartheta]$, the system-bath-coupling strength $\alpha(k)$ and $\beta(k)$, which are proportional to $k^{1/2}$, are approximately constant and we set them to $\alpha(k)\rightarrow \alpha({k_0}) \equiv \alpha$ and $\beta(k)\rightarrow \beta({k_0}) \equiv \beta$. Thus the $k$-integrals should be understood as integrals of the form
\begin{equation}
\int_{0(-\infty)}^\infty \mathrm{d}k \;\alpha(k)\ldots \longrightarrow \alpha(k_0)\int_{k_0-\vartheta/c}^{k_0+\vartheta/c} \mathrm{d}k\ldots.
\end{equation}
This is the first Markov approximation. Formally integrating equations \eqref{eq:diffcAt} and \eqref{eq:diffcBt} and inserting them in the first one yields terms of the form $f_\pm(t,t')\equiv e^{i k[\ell(t)\pm\ell(t')]}$. When the cutoff frequency $\vartheta$ is chosen in a way that $\vartheta\ell_0\ll c$ we can set $\ell_0 k=\ell_0 k_0$ so that they then become independent of $k$. The remaining $k$-integrals over this frequency band give rise to nascent delta functions of the form
\begin{equation*}
\int_{k_0-\vartheta/c}^{k_0+\vartheta/c} \mathrm{d}k\,  e^{i(\omega(k)-\omega_0)(t'-t)}\equiv\frac{2\pi}{c}\delta_\vartheta(t'-t)
\end{equation*}
and we find
\begin{equation*}
\begin{split}
\dot{\tilde{c}}(t)=\frac{\pi g^2}{2c}\int_0^t \mathrm{d}t'\;c(t')&\left\{f_+^*(t,t')\delta_\vartheta(t'-t+\tau)e^{i\omega_0\tau}\right.\\
&-[f_-(t,t')+f_-^*(t,t')]\delta_\vartheta(t'-t)\\
&\left.+f_+(t,t')\delta_\vartheta(t'-t-\tau)e^{-i\omega_0\tau}\right\}\\
-\frac{\pi h^2}{c}\int_0^t \mathrm{d}t'\;c(t')&\delta_\vartheta(t'-t).
\end{split}
\end{equation*}
Since the amplitude, $\tilde c(t)$, in a rotating frame varies on a timescale approximately given by $\gamma$ and the mirror on a timescale $k_0\ell_0\nu$ we can take the limit $\vartheta\rightarrow\infty$, if we assume $\max{\{\gamma,k_0\ell_0\nu\}}\ll\vartheta$, because all optical frequencies have been transformed away. In this limit we can use the nascent delta functions like arbitrary delta functions and obtain (\ref{eq:nonmarkovpop}). This approximation scheme is independent of the chosen cutoff $\vartheta$. Summarizing, we find that the cutoff frequency $\vartheta$ has to fulfill
\begin{equation*}
\max{\{\gamma,k_0\ell_0\nu\}}\ll\vartheta\ll\min{\{c/\ell_0, \omega_0\}}.
\end{equation*}
Expanding $f_+^*(t,t-\tau)$ with the Jacobi-Anger identity \cite{jacobianger}
\begin{equation*}
f_+^*(t,t-\tau)=\sum_{k=-\infty}^{\infty}\Pi_k(k_0\ell_0,\nu\tau)\;e^{-ik\nu t},
\end{equation*}
where $\Pi_k(k_0\ell_0,\nu\tau)$ is defined in \eqref{Pifactor}, and Laplace transform of Eq.~\eqref{eq:nonmarkovpop} yields
\begin{equation*}
s\hat c(s)-c(0)=-\frac{\gamma}{2}\hat c(s)+\epsilon\frac{\gamma}{2}\sum_{k=-\infty}^{\infty}\Pi_ke^{i(\omega_0-k\nu)\tau}e^{-s\tau}\hat c(s+ik\nu).
\end{equation*}
Iterating in $\epsilon$ and transforming back yields \eqref{excitedstate}.

\section{Derivation of the Spectrum}\label{app:wwspectrum}
In this appendix we calculate the equation for the electric field,  Eq.~\eqref{cField}, and the equation for the emission spectrum, Eq.~\eqref{eq:spectrum}.
Inserting (\ref{multimode}) in the c-number electric field (\ref{cFieldformal}) we find integrals of the form
\begin{multline*}
\int_0^\infty d\omega \;\mathrm{J}_n(k\ell_0)\;\mathrm{J}_m(k\ell_0)\;\frac{e^{i\omega\tau}}{i(\omega-\tilde\omega_0-n\nu)-\tilde\gamma/2}\\
\approx\mathrm{J}_n(k_0\ell_0)\;\mathrm{J}_m(k_0\ell_0)\;\int_{-\infty}^\infty d\omega \;\frac{e^{i\omega\tau}}{i(\omega-\tilde\omega_0-n\nu)-\tilde\gamma/2}\\
=-2\pi\,\mathrm{J}_n(k_0\ell_0)\;\mathrm{J}_m(k_0\ell_0)\;\Theta(-\tau)\;e^{(i\tilde\omega_0+in\nu+\tilde\gamma/2)\tau}.
\end{multline*}
Using $\sum_m\;\mathrm{J}_{k+m}(k_0\ell_0)\;\mathrm{J}_m(k_0\ell_0)=\delta_{k0}$ we find expression (\ref{cField}).

Filtering this field yields with (\ref{cField})
\begin{multline*}
w(t,\Gamma_D,\omega_D)
=\Big|\Gamma_D\frac{\alpha g\pi}{2c}\;\left(\mathcal{L}(t)\right.\\
\left.-\sum_{n,m}\;\mathrm{J}_n(k_0\ell_0)\;\mathrm{J}_m(k_0\ell_0)\mathcal{L}_{nm}(t)\right)\;\Big|^2.
\end{multline*}
In the limit of ideal spectral resolution $\Gamma_D\longrightarrow 0$ and in the Markovian limit $t\gg1/\tilde\gamma, R/c, d/c$ we find
\begin{equation}
\begin{split}
\mathcal{L}(t)&=-\frac{e^{-i \omega_D(t-\frac{d}{c})} }{i(\omega_D-\tilde\omega_0)-\tilde\gamma/2},
\end{split}
\end{equation}
and
\begin{equation}
\begin{split}
\mathcal{L}_{nm}(t)&=-\frac{e^{-i \omega_D(t-\frac{2R+d}{c})}e^{-i(n+m)\nu \frac{2R+d}{c}}}{i(\omega_D-\tilde\omega_0-(n+m)\nu)-\tilde\gamma/2}.
\end{split}
\end{equation}
Identifying the spectrum as the normalized energy density for different resonant frequencies of the cavity we result in (\ref{eq:spectrum}).


\begin{thebibliography}{10}
\bibitem{blatt}  Eschner J.,  Raab C.,  Schmidt-Kaler F., and R. Blatt, \textit{Nature} (London)  \textbf{413}, 495 (2001)
\bibitem{blatt2}M. A. Wilson, P. Bushev, J. Eschner, F. Schmidt-Kaler, C. Becher, and R. Blatt and U. Dorner \textit{Phys. Rev. Lett.} \textbf{91}, 213602 (2003)
\bibitem{Dubin2007} F. Dubin, D. Rotter, M. Mukherjee, C. Russo, J. Eschner, R. Blatt, Phys. Rev. Lett. \textbf{98} 183003 (2007)

\bibitem{Morawitz} H. Morawitz, Phys. Rev. {\bf 187}, 1792 - 1796 (1969)
\bibitem{Cook1987} R.J. Cook, P. W. Milonni, Phys. Rev. A {\bf 35}, 5081 (1987).
\bibitem{Alber1992} G. Alber, Phys. Rev. A {\bf 46}, R5338-R5341 (1992).



\bibitem{Dorner} Dorner U., Zoller P., \textit{Phys. Rev.} A, \textbf{66} 023816 (2002)
\bibitem{Viktor} P. Bushev, D. Rotter, A. Wilson, F. Dubin, Ch. Becher, J. Eschner, R. Blatt, V. Steixner, P. Rabl, and P. Zoller \textit{Phys. Rev. Lett.} \textbf{96}, 043003 (2006)

\bibitem{Horak2009} P. Horak, A. Xuereb, T. Freegarde, arXiv:0904.3059
\bibitem{Xuereb2009a} A. Xuereb, P. Horak, T. Freegarde, Phys. Rev. A \textbf{80}, 013836 (2009)
\bibitem{Xuereb2009b} A. Xuereb, P. Domokos, J. Asboth, P. Horak, T. Freegarde, Phys. Rev. A \textbf{79}, 053810 (2009)

\bibitem{Marquardt2009} F. Marquardt, S.M. Girvin, arXiv:09050566 (2009).
\bibitem{Eberly1} J. H. Eberly, K. Wódkiewicz, \textit{J. Opt. Soc. Am.} \textbf{67}, 1252 (1977)
\bibitem{Eberly2}  J. H. Eberly, C. V. Kunasz and K. W\'odkiewicz, \textit{ J. Phys. B: At. Mol. Phys.} \textbf{13} 217 (1980)


\bibitem{milonni} Milonni P. W. "\textit{Quantum Vacuum - An Introduction to QED}" Academic Press (1993)
\bibitem{jacobianger} M. Abramowitz and I. A. Stegun,  \textit{Handbook of Mathematical Functions},  p. 361, 9.1.42 - 45 (1965)

\bibitem{colgard} M. J. Collett and C. W. Gardner, \textit{Phys. Rev.}, A \textbf{30}, 1386 (1984)




\end{thebibliography}
\end{document}